\documentclass[aps,prl,preprintnumbers,groupedaddress,showpacs,twocolumn,floatfix]{revtex4}

\usepackage{graphicx}

\begin{document}

\newcommand{\bfm}[1]{\mbox{\boldmath$#1$}}
\newcommand{\bff}[1]{\mbox{\scriptsize\boldmath${#1}$}}

\title{Measuring  Vacuum Polarization with Josephson
Junctions}

\author{Alexander A.~Penin}
\email{apenin@phys.ualberta.ca}
\affiliation{Department of Physics, University of Alberta,
Edmonton, AB T6G 2J1, Canada}
\affiliation{
Institute for Nuclear Research of Russian Academy of Sciences,
117312 Moscow, Russia}


\begin{abstract}
We argue that the vacuum polarization by the virtual electron-positron pairs can
be measured by studying a Josephson junction in a strong magnetic field.
The vacuum polarization results in a  weak dependence of the Josephson constant
on the magnetic field strength which is within the reach of the existing
experimental techniques.
\end{abstract}


\maketitle


Interaction of the electromagnetic field with the quantum fluctuations of vacuum
is responsible for  many remarkable nonlinear phenomena described by the theory
of quantum electrodynamics (QED) \cite{HeiEul,Sch}. In particular, the
polarization of vacuum by virtual electron-positron pairs results in dependence
of a charged particle interaction with the electromagnetic potential on the
field strength and on the characteristic momentum transfer. This intrinsically
relativistic effect is strongly suppressed and becomes observable only in very
strong fields or in high energy processes. For example, in electron-positron
scattering at  energy close to the mass of the $Z$-boson, which is five orders
of magnitude heavier than an electron, the electromagnetic interaction is
described by the ``running'' coupling constant $\alpha(M_Z)\approx 1/128$ rather
than by the usual fine structure constant $\alpha\approx 1/137$. In condensed
matter the effect of vacuum polarization is so tiny that in most cases it is
absolutely indistinguishable against the background of the complex quantum
mechanical interactions. The only chance to observe it is to find a relation
which is  exact in quantum mechanics but can be modified in full QED. A renowned
example of an exact result in quantum mechanics is the Josephson
frequency-voltage relation.  In the seminal papers \cite{Jos} Josephson studied
a system of  two superconductors separated by a thin insulating barrier, the
Josephson junction. He found, in particular, that a {\it constant} voltage $V$
across the junction results in an {\it alternating} current through the junction
at the frequency $\nu$ proportional to the voltage, $\nu=K_JV$, where $K_J$ is
the Josephson constant. A simple quantum mechanical calculation relates it to
the electron charge and the Planck constant 
\begin{equation}
K_J=2e/h. 
\label{jk}
\end{equation}
This prediction has been verified experimentally \cite{Sha} and is known as ac
Josephson effect. A remarkable property of the ac Josephson effect is that 
Eq.~(\ref{jk}) is stable against all kinds of perturbations in quantum mechanics
because of the gauge invariance \cite{Blo}.  Another famous example of the exact
relations in quantum mechanics is the quantization of the  Hall conductivity  of
the two-dimensional electron system in a strong transverse magnetic field  in
the units of $1/R_K$, where 
\begin{equation}
R_K=h/e^2
\label{rk}
\end{equation}
is the von Klitzing constant \cite{KDP}. As in the case of the ac Josephson
effect, this result is protected against corrections by the  gauge invariance
\cite{Lau1}. The absence of quantum mechanical corrections makes
Eqs.~(\ref{jk},\ref{rk}) crucial for metrology. For example, the most accurate
value of the Planck constant is currently obtained through the relation
$1/h=K_J^2R_K/4$ \cite{MohTay}. Given the important role  that $K_J$ and $R_K$
play for determination of the fundamental constants, much effort is being made
to verify Eqs.~(\ref{jk},\ref{rk}) experimentally \cite{Kel}. 

On the theory side, for a long time these relations were thought to be exact.
Recently, however, a deviation from the quantum mechanical result~(\ref{rk}) has
been discovered \cite{Pen}. The physics behind this phenomenon is in a
modification of the {\it local} coupling of charged particles to the
electromagnetic potential due to vacuum polarization by highly virtual
electron-positron pairs in a  strong magnetic field.  For a typical magnetic
field strength of about $10$~T it amounts to a tiny $10^{-20}$ correction. This
is  well beyond the precision of the current quantum Hall experiments
\cite{SchPoi}, which is about one part in $10^{12}$ and is limited by the
thermal Johnson-Nyquist noise. For Josephson junctions there is no such 
limit  and universality of the frequency-voltage relation has been established
to the amazing accuracy of $10^{-19}$ already two decades ago \cite{TJL}. Thus
one may expect that a similar effect, if it exists, can be experimentally
observed in a Josephson junction subject to a strong magnetic field. The
purpose of this letter is to show that this is indeed the case.

Our analysis of the ac Josephson effect is  based on the following fundamental
properties: (i) existence in a superconductor of the macroscopic phase-coherent
state of  weakly bound electron pairs (Cooper pairs) described by the wave
function $\Psi=|\Psi|e^{i\theta}$, (ii) $2\pi$-periodic dependence of the
current through the junction on the difference of the phase $\theta$ across the
junction, and  (iii) gauge invariance of the electromagnetic interactions. We
also rely on  the fact  that a sufficiently strong magnetic field is not
screened by the Josephson current and penetrates the junction  \cite{OweSca}. To
get the correction to Eq.~(\ref{jk}) in a closed analytical form we consider a
simplified model of the Josephson junction described below. This, however,  does
not affect the general character of the result.

Interaction of Cooper pairs with the external electromagnetic potential
$A^\mu=(A_0, {\bfm A})$ is dictated by gauge invariance and in quantum
mechanics is described by the Hamiltonian of the following general 
form \cite{footnote}
\begin{equation}
{\cal H}=2eA_0+H(\bfm{B},\bfm{E},\bfm{D}),
\label{ham} 
\end{equation}
where the second term is a function of the spatial covariant derivative ${\bfm
D}={\bfm \partial}-i2e{\bfm A}$, the electric field $\bfm{E}$ and the magnetic
field $\bfm{B}$ corresponding to the potential $A^\mu$. Our analysis does not
depend on the specific form of the function $H$. The gauge invariance ensures
that Eq.~(\ref{ham}) is linear in  $A_0$ and explicitly depends on ${\bfm A}$ 
only through the covariant derivative. The form of the interaction~(\ref{ham}) 
guarantees that the  Josephson frequency-voltage relation is exact 
\cite{Blo}. This can be proven as follows. The physical observables depend
on the gauge invariant combination 
\begin{equation}
\phi(\bfm{r})=\theta(\bfm{r})-2e\int^{\bff{r}}\bfm{A}(\bfm{r}')\cdot
d\bfm{r}'
\label{phi}
\end{equation}
rather than on the bare phase  $\theta(\bfm{r})$. At the same time the  static 
scalar potential $A_0(\bfm{r})$ can be the removed by the  gauge transformation
\begin{equation}
{A^\mu}'(r)=A^\mu(r)-\partial^\mu\xi(r), \qquad
\xi(r)=tA_0(\bfm{r}),
\label{xi}
\end{equation}
so that
\begin{equation}
A_0'(\bfm{r})=0, \qquad \bfm{A}'(r)=\bfm{A}(\bfm{r})+t\bfm{\nabla}A_0(\bfm{r}).
\label{gauge}
\end{equation}
Thus the phase difference between two arbitrary points $\bfm{r}_1$ and
$\bfm{r}_2$ can be written as
\begin{eqnarray}
\Delta\phi&=&\Delta\phi_0-2et
\int^{\bff{r}_1}_{\bff{r}_2}\bfm{\nabla}A_0(\bfm{r}')\cdot d\bfm {r}'
\nonumber \\
&=&
\Delta\phi_0-2et\left(A_0(\bfm{r}_1)-A_0(\bfm{r}_2)\right),
\label{delphi}
\end{eqnarray}
where $\Delta\phi_0$ is the time-independent phase difference in the absence of
the electric field. The  Josephson current is $2\pi$-periodic function of $\phi$
and, therefore,  Eq.~(\ref{delphi}) implies that the potential difference
$V=A_0(\bfm{r}_1)-A_0(\bfm{r}_2)$ between two superconductors results in the
Josephson current oscillations with the angular
frequency
\begin{equation}
\omega=2eV,
\label{omega}
\end{equation}
which gives Eq.~(\ref{jk}) in the physical units. Thus  gauge invariance leaves
no room for corrections to this equation in quantum mechanics.

\begin{figure}[t]
\begin{center}
\hspace*{-10mm}
\begin{picture}(0,0)
\includegraphics[width=7.5cm]{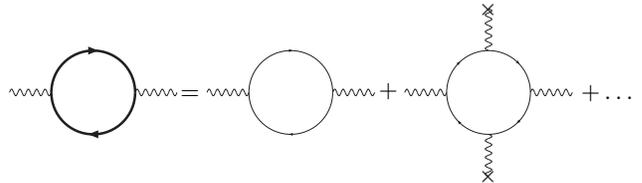}
\end{picture}
\begin{picture}(220,110)
\put(233,33.1){\makebox(0,0)[r]{$+\ldots$}}
\put(69,30.9){\makebox(0,0)[rb]{$=$}}
\put(144,30.9){\makebox(0,0)[rb]{$+$}}
\end{picture}
\end{center}

\caption{\small{Feynman diagrams representing the  vacuum polarization
by virtual electron-positron pairs in the external magnetic field.
The arrow and wavy lines correspond to  the free electron and photon
propagators, respectively. The bold arrow lines correspond to the electron
propagating in the external magnetic field. The crossed wavy lines represent 
the external magnetic field.} \label{fig1}}
\end{figure}

In full QED  the situation becomes more involved since the  coupling of charged
states to the electromagnetic potential  is modified by the radiative
corrections though it remains gauge invariant. We are interested in the
corrections due to vacuum polarization through creation and annihilation of
virtual electron-positron pairs in external magnetic field  graphically
shown  in Fig.~\ref{fig1}. Quantitatively the effect is determined by the
behavior of the vacuum polarization tensor $\Pi_{\mu\nu}(q)$ at small
four-momentum transfer $q$. It can be expanded in powers of the external field,
see Fig.~\ref{fig1}. For a homogeneous or slowly varying field this gives a
series in the parameter
\begin{equation}
\beta^2 =\left(\frac{eB}{m^2}\right)^2\ll 1, 
\label{beta}
\end{equation} 
where $B=|\bfm{B}|$ and $m$ is the electron mass. The leading ${\cal O}(1)$ term
of the expansion is ultraviolet divergent and is absorbed by the on-shell
renormalization of the physical electron charge $e$. The ${\cal O}(\beta^2)$ 
correction  to the polarization tensor in the limit $q\to 0$ reads \cite{Pen}
\begin{eqnarray}
\lefteqn{\delta\Pi_{\mu\nu}(q)=-\frac{\alpha}{\pi}
\beta^2\frac{1}{45}\bigg[ 2\left(g_{\mu\nu}q^2-q_\mu q_\nu\right)}
\nonumber\\
&&-7\left(g_{\mu\nu}q^2-q_\mu q_\nu\right)_\parallel
+4\left(g_{\mu\nu}q^2-q_\mu q_\nu\right)_\perp\bigg].
\label{delpi}
\end{eqnarray}
The correction to the polarization tensor is transverse  because of gauge
invariance. At the same time the Lorentz and rotational invariance  is broken
and Eq.~({\ref{delpi}}) includes the transverse projectors in the ``parallel''
$(q_0,\bfm{q}_\parallel)$ and ``orthogonal'' $(\bfm{q}_\perp)$ two-dimensional
subspaces of the whole four-dimensional Minkowskian momentum space
$(q_0,\bfm{q})$.  Here $\bfm{q}_\parallel$ and $\bfm{q}_\perp$ components
correspond to the spatial momentum parallel and orthogonal to the magnetic
field, respectively.  The correction~(\ref{delpi}) cannot  be ``renormalized
out''  and results in corrections to the local coupling of charged states to
the electromagnetic potential and the correction to the photon propagator, which
is nontrivial since the external magnetic field changes the photon dispersion
law \cite{Adl}. The first  term of Eq.~({\ref{delpi}}) is Lorentz
covariant and has the same structure as the ${\cal O}(1)$ vacuum polarization.
The last two terms of Eq.~({\ref{delpi}}) violate Lorentz and  rotational
invariance leaving unbroken axial symmetry in respect to the magnetic field
direction. These noncovariant terms result in the double pole
contribution in the photon propagator and contribute to the
photon dispersion.

\begin{figure}[t]
\begin{center}
\hspace*{-15mm}
\begin{picture}(0,0)
\includegraphics[width=3.8cm]{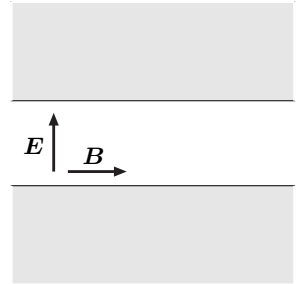}
\end{picture}
\begin{picture}(70,110)
\put(10,53){\makebox(0,0)[r]{$\bfm E$}}
\put(33,49){\makebox(0,0)[r]{$\bfm B$}}
\end{picture}
\end{center}

\caption{\small{Model geometry of the Josephson junction. Gray
areas correspond to  the superconductors separated by the
insulator layer. The electric and magnetic fields are
homogeneous in the insulator and vanish in the superconductors.} \label{fig2}}
\end{figure}

We are interested only in the corrections to the local
coupling with the scalar potential $A_0$, which depends on the magnetic field
configuration, {\it i.e.} on the particular structure of the Josephson
junction. To get this correction in a closed form we consider a simplified model
of the Josephson junction shown in Fig.~\ref{fig2}. It corresponds to the
homogeneous orthogonal electric and magnetic fields inside the infinite and
plane insulator layer.  For such a field configuration the analysis is 
simplified because the noncovariant terms vanish and the correction to the
dispersion law does not emerge. Thus the  correction to the Hamiltonian takes
the standard QED form
\begin{equation}
\delta_{v.p.}{\cal
H}=-e\int
\frac{\delta\Pi_{0\nu}(q)}{q^2}\tilde{A}^\nu(q)
e^{iqr}\frac{d^4q}{(2\pi)^4}+\ldots,
\label{delham}
\end{equation}
where  $\tilde{A}^\nu(q)$ is the Fourier transform of the electromagnetic
potential and we keep only the contribution with the structure of the first term
in  Eq.~({\ref{ham}}). The integral in Eq.~({\ref{delham}}) gets
nonvanishing contribution only from the first
term of Eq.~({\ref{delpi}}) and can be evaluated with the result
\begin{equation}
\delta_{v.p.}{\cal H}=2\delta e A_0(\bfm{r}),
\label{delres}
\end{equation}
where  $A_0(\bfm{r})=\bfm{E}\cdot\bfm{r}$ and 
\begin{equation}
\delta e=\frac{1}{45}\frac{\alpha}{\pi} \beta^2e.
\label{dele}
\end{equation}
Eq.~({\ref{delres}}) has exactly the same form as the first term of
Eq.~({\ref{ham}}) but it is {\em gauge invariant} because $\delta\Pi_{\mu\nu}$
in Eq.~({\ref{delham}}) is transverse. It is easy to check by explicit
calculation that Eq.~({\ref{delres}}) does not change under the gauge
transformation~(\ref{gauge}). This can be understood as a manifestation of
the Ward indentity \cite{War}, which means that the interaction of the charged 
states to the pure gauge field configurations is not renormalized.
However this remaining ``potential'' term can be removed from the Hamiltonian by
an additional gauge transformation with the parameter
\begin{equation}
\delta\xi(r)=\frac{\delta e}{e}\xi(r).
\label{delxi}
\end{equation}
Then, by using the same arguments as before, we derive the new result for the
Josephson current frequency
\begin{equation}
\omega=2e^*V,
\label{omeganew}
\end{equation}
which differs from Eq.~(\ref{omega}) by the ``effective charge'' $e^*=e+\delta
e$. This is not a surprising result because $2e^*V$ is nothing 
but the difference $\Delta\mu$ of the electrochemical potential between
two identical superconductors and the  general form of the Josephson
relation
\begin{equation}
\omega=\Delta\mu
\label{omegamu}
\end{equation}
is not changed. 

The vacuum polarization effect can be accounted for by
introducing an effective field-dependent Josephson constant
\begin{equation}
K_J(B)=K_J\left[1+\frac{1}{45}\frac{\alpha}{\pi}\beta^2\right],
\label{res}
\end{equation}
or in  physical units
\begin{equation}
K_J(B)=K_J\left[1+\frac{1}{45}\frac{\alpha}{\pi}
\left(\frac{\hbar eB}{c^2m^2}\right)^2\right]. 
\label{ressi}
\end{equation}
This result is not exact since it was obtained for a simplified model of the
junction. In reality the magnetic field in the junction is partially screened by
the Josephson current and oscillates about some average value \cite{OweSca},
which should be used in Eq.~(\ref{ressi}). At the same time the result is not
affected by the sharp variation of the magnetic field near the superconductor 
surface because $A_0$ is continuous and the thin boundary region does not
contribute to the potential difference. In general, the frequency-voltage
relation is not sensitive to the interaction inside the superconductors where
$A_0$ is constant. 

A more subtle problem  is that in real experiments the magnetic field does not
vanish only in a finite volume. One may argue that the electric charge of a
Cooper pair measured by means of the Gauss law at spatial infinity in this case
does not differ from $2e$, that is in apparent contradiction with
Eq.~(\ref{omeganew}).   Similar argument has been used in
Refs.~\cite{LanSch,HSS} to prove the absence of the corrections to $K_J$ through
the electron coupling modification due to the interaction of  the  electrons
inside  the superconductor. In our case, however, this argument does not work
since the coupling is modified outside the superconductor, where the scalar
potential varies. Indeed, the chemical potential difference between two
superconductors  is given by  $\Delta\mu=2e^*V$ if the magnetic field is
homogeneous in the region of the non-vanishing electric field giving rise to 
$V$, {\em i.e.} in the vicinity of the junction, regardless to its behavior at
the infinity. Here the following analogy may be useful: the vacuum polarization
in the interatomic electric and magnetic fields does not change the total charge
of the atom but does change the electron binding energy. 

Let us now examine  prospects to  detect the magnetic field dependence of $K_J$
experimentally. A relevant  technique has been  elaborated long time ago and
consists in comparison of the voltage difference between two junctions which are
phase locked to a source of microwave radiation \cite{Cla}.  The junctions are
connected by the superconducting links to form a loop.  A nonvanishing voltage
difference results in a loop current increasing linearly in time, which can be
monitored by a sensitive SQUID  detector. The time-independent magnetic field
does not change the frequency of the  plane waves, though it changes their
dispersion \cite{Adl}. Thus if one of the junctions is embedded into
magnetic field, the correction to the frequency-voltage  relation results in a
net electromotive force around the loop. Note that the junctions themselves do
not contribute to the net electromotive force because of Eq.~(\ref{omegamu}). To
estimate the size of the effect we rewrite the correction term of
Eq.~(\ref{ressi}) as follows
\begin{equation}
\frac{1}{45}\frac{\alpha}{\pi}\left(\frac{B}{B_0}\right)^2,
\label{cor}
\end{equation}
where $B_0=c^2m^2/(\hbar e)\approx 4.41\cdot 10^{9}$~T. The critical magnetic
field which destroys the quantum coherence in the Josephson junction is close to
the one of the superconductors and could be as large as a few units times
$10$~T. This gives approximately $10^{-20}$ variation of the Josephson constant.
On the other hand a pair of junctions had been compared with relative accuracy
of $10^{-19}$ to test the equivalence principle for charged particles \cite{TJL}
and  the accuracy can probably further increased. Thus the effect is likely to
be within the reach of the existing experimental techniques.


In summary, the vacuum polarization  alters the Josephson frequency-voltage
relation  in the presence of a strong magnetic field and  results in a weak
dependence of the Josephson constant on the magnetic field strength. This 
remarkable  manifestation of a fine nonlinear quantum field effect in a
collective phenomenon in condensed matter could be observed in a dedicated
experiment, that would literally be a measurement of the vacuum polarization
with a voltmeter.

\begin{acknowledgements}
I would like to thank W. Poirier for useful communications.
I am grateful to  K. Melnikov for the  discussions and cross-checks.
This work  is supported  by the Alberta Ingenuity foundation
and NSERC.
\end{acknowledgements}



\begin{thebibliography}{99}

\bibitem{HeiEul} W. Heisenberg and H. Euler, Z. Physik {\bf 98}, 714 (1936).

\bibitem{Sch} J. Schwinger, Phys. Rev.  {\bf 82}, 664 (1951).

\bibitem{Jos} B.D. Josephson, Phys. Lett. {\bf 1}, 251 (1962); Adv. Phys. {\bf
14}, 419 (1965).

\bibitem{Sha}   S. Shapiro, Phys. Rev. Lett. {\bf 11}, 80 (1963).


\bibitem{Blo} F. Bloch, Phys. Rev. Lett. {\bf 21}, 1241 (1968); Phys. Rev. B
{\bf 2}, 109 (1970).

\bibitem{KDP} K. von Klitzing, G. Dorda, and M. Pepper, Phys. Rev. Lett. {\bf
45}, 494 (1980).

\bibitem{Lau1} R.B. Laughlin, Phys. Rev. B {\bf 23}, 5632 (1981).


\bibitem{MohTay} P.J. Mohr and  B.N. Taylor, Rev. Mod. Phys. {\bf 77}, 1 (2005).

\bibitem{Kel} M.W. Keller, Metrologia {\bf 45} 102 (2008).

\bibitem{Pen} A. Penin, Phys. Rev. B {\bf 79}, 113303  (2009); Erratum {\it
ibid.}  {\bf 81}, 089902(E) (2010).


\bibitem{SchPoi} F. Schopfer and W. Poirier, J. Appl. Phys. {\bf 102}, 054903
(2007).

\bibitem{TJL} A.K. Jain, and J.E. Lukens, and J.-S. Tsai, Phys. Rev. Lett. {\bf
58}, 1165 (1987).


\bibitem{OweSca} C. S. Owen and D. J. Scalapino, Phys. Rev. {\bf 164}, 538
(1967).

\bibitem{footnote} In the rest of the paper, if
it is not explicitly stated otherwise, we adopt the system of units
used in particle physics, where $\hbar=c=1$ and
$\alpha=e^2/(4\pi)$.

\bibitem{Adl} S.L. Adler, Ann. Phys. (N.Y.) {\bf 67}, 599 (1971).

\bibitem{War}   J.C. Ward,  Phys.Rev. {\bf 78}, 182 (1950).


\bibitem{LanSch} D.N. Langenberg and J.R. Schrieffer, Phys. Rev. B {\bf 3},
1776 (1971).

\bibitem{HSS} J.B. Hartle, D.J. Scalapino, and R.L. Sugar, Phys. Rev. B {\bf 3},
1778 (1971).

\bibitem{Cla} J. Clarke, Phys. Rev. Lett. {\bf 21}, 1566 (1968)














\end{thebibliography}
\end{document}